\documentclass[preprintnumbers,elsart,showpacs]{revtex4}
\usepackage{amssymb}
\usepackage{amsmath}
\usepackage{graphicx}
\usepackage{dcolumn}
\usepackage[center]{subfigure}
\usepackage{float}
\usepackage{color}

\begin{document}

\title{Self-trapping under the two-dimensional spin-orbit-coupling and
spatially growing repulsive nonlinearity}
\author{Rongxuan Zhong$^{1}$, Zhaopin Chen$^{2}$, Chunqing Huang$^{1}$,
Zhihuan Luo$^{3}$, Haishu Tan$^{1}$, Boris A. Malomed$^{2,4,1}$, Yongyao Li$%
^{1}$}
\email{yongyaoli@gmail.com}
\affiliation{$^{1}$School of Physics and Optoelectronic Engineering, Foshan University,
Foshan 528000, China \\
$^{2}$ Department of Physical Electronics, School of Electrical Engineering,
Faculty of Engineering, Tel Aviv University, Tel Aviv 69978, Israel.\\
$^{3}$College of Electronic Engineering, South China Agricultural
University, Guangzhou 510642, China\\
$^{4}$ITMO University, St. Petersburg 197101, Russia }

\begin{abstract}
We elaborate a method for the creation of two- and one-dimensional (2D and
1D) self-trapped modes in binary spin-orbit (SO)-coupled Bose-Einstein
condensates (BECs) with the contact repulsive interaction, whose local
strength grows fast enough from the center to periphery. In particular, an
exact semi-vortex (SV) solution is found for the anti-Gaussian
radial-modulation profile. The exact modes are included in the numerically
produced family of SV solitons. Other families, in the form of mixed modes
(MMs), as well as excited state of SVs and MMs, are produced too. While the
excited states are unstable in all previously studied models, they are
partially stable in the present one. In the 1D version of the system, exact
solutions for the counterpart of the SVs, namely, \textit{semi-dipole}
solitons, are found too. Families of semi-dipoles, as well as the 1D version
of MMs, are produced numerically.\newline
\end{abstract}

\maketitle

\section{Introduction}

For a long time, prediction and experimental creation of stable two- and
three dimensional (2D and 3D) solitons remains a challenging problem of
nonlinear physics. A commonly known difficulty is that fundamental
multidimensional solitary modes in 2D and 3D media with the ubiquitous cubic
self-focusing nonlinearity are made unstable, respectively, by the critical
and supercritical collapse \cite{Berge1998,Fibich1999} in the 2D and 3D
geometry. Solitons with embedded vorticity (alias vortex rings and tori)
\cite{Desyatnikov2005} are destabilized by a still stronger splitting
azimuthal instability. Elaboration of methods for stabilization of
multidimensional fundamental and vortex solitons is an issue of great
interest to nonlinear photonics and mean-field dynamics in Bose-Einstein
condensates (BECs)\cite{Review2005}-\cite{Jianhua}.

A method which makes it possible to produce exceptionally robust
multidimensional modes is based on using the repulsive (defocusing)
nonlinearity, whose local strength grows from the center to periphery, in
the space of dimension $D$, at any rate faster than $R^{D}$, as demonstrated
in a number of theoretical works \cite{Borovkova2011}-\cite{Driben20142}.
This type of the nonlinearity modulation can be induced by means of various
techniques. In optical media, one may use nonlinearity-enhancing dopants
with an inhomogeneous density \cite{Hukriede2003}. In BEC, the tunability of
the local nonlinearity by means of the magnetic Feshbach resonance (FR) \cite%
{Inguscio,Pollack} suggests a possibility for the creation of spatially
modulated nonlinearity profiles by means of appropriately shaped magnetic
fields \cite{14Abdullaev}, which was realized in the experiment \cite{Clark}%
. Furthermore, nearly arbitrary spatial profiles of the self-repulsive
nonlinearity can be imposed by means of the optically controlled FR \cite%
{Yamazaki,15Yan2013}, as well as with the help of combined magneto-optical
shaping \cite{16bauer2009}. A nonlocal version of this setting may be
realized in terms of the long-range interaction between dipolar moments
locally induced by a nonuniform dc electric field \cite%
{17Yongyao2013,18FKh2014,YLiFOP}.

The other promising method for the stabilization of 2D and 3D solitons
relies upon the use of the effective spin-orbit (SO) coupling in binary
condensates. The possibility to emulate effects of the SO coupling,
originally known in the physics of semiconductors, in two-component atomic
BEC has been demonstrated experimentally and analyzed in detail
theoretically \cite{YJLIN2011}-\cite{Yongping2016}. The SO coupling in BEC
being a linear interaction between spatially inhomogeneous states in the two
components, its interplay with the intrinsic nonlinearity of the BEC makes
it possible to predict diverse nonlinear phenomena \cite{CWang2010}-\cite%
{Yongyao2017}. In particular, a noteworthy result, which opens the way to
novel applications of the SO coupling in the studies of matter waves, is the
stabilization of 2D solitons in the free space, in the form of semi-vortices
(SVs) (alias half-vortices \cite{Drummond}), with the vorticity carried by
one component of the spinor complex, and mixed-modes (MMs), which combine
terms with zero and nonzero vorticities in each component \cite%
{SVS1,SVS2,Guihua2017}. These results break a commonly adopted paradigm
stating that bright solitons supported by cubic self-attractive
nonlinearities in the free space are always unstable, due to the presence of
the critical collapse in the same system \cite{Berge1998,Fibich1999}. In
other words, the stable SVs and MMs play the role of the ground state, which
is missing in the 2D free-space settings with the cubic self-attraction, in
the absence of the SO coupling \cite{SVS1,SVS2}.

In optics, it has been demonstrated that spatiotemporal solitons in a planar
dual-core Kerr-nonlinear waveguide can be stabilized by temporal dispersion
of the linear coupling between the cores, which provides for an optical
counterpart of the SO coupling \cite{YVK20132}
. Further, as concerns emulation of the SO coupling in photonics, interest
was recently drawn to effects in exciton-polariton condensates in
microcavities \cite{Bardyn2015}-\cite{SMS}.

In the 3D setting, it was found that the interplay of the SO coupling with
the attractive cubic nonlinearity in the free space creates metastable 3D
solitons of the same types as in 2D, \textit{viz}., SVs and MMs \cite{3DSOC}%
. The SO coupling also helps one to build stable solitons under the action
of long-range dipole-dipole interactions in binary BECs. As reported in
Refs. \cite{Xuyong2015,Xunda2016,Bingjin2017}, these may be striped modes
and anisotropic vortices, as well as stable 2D gap solitons, supported by
the combination of the SO coupling and Zeeman splitting, in the case when
the kinetic-energy terms may be neglected in the respective 2D system \cite%
{Yongyao2017,gapsoliton2}. Peculiarities of the collapse in the SO-coupled
system were explored in Refs. \cite{Dias2015} and \cite{Sherman}.

The objective of this work is to study solitons in the setting which was not
explored before, namely, the SO coupling acting in a combination with
self-repulsive contact nonlinearity, which induces self-trapping due to the
growth of its local strength from the center to periphery, similar to what
was introduced (in the absence of the SO-coupling) in Refs. \cite%
{Borovkova2011} and \cite{Borovkova20112}. In particular, we find stable
analytical solutions for 2D solitons of the SV type, which, to the best of
our knowledge, is the first example of solitons available in an exact form
under the action of the SO coupling. Stable 2D solitons of the MM type, as
well as unstable excited states of SVs and MMs, are found in a numerically
form. Exact solitons solutions are also reported for the 1D version of the
spinor system.

The paper is structured as follows. The model is introduced in Section II,
which is followed by the presentation of the analytical and numerical
solutions for SV solitons and results for their stability in Section III.
Solitons of the MM type are addressed by means of numerical methods in
Section IV. Further, excited states of the SVs and MMs are considered in
Section V. Exact analytical solutions and families of numerically found ones
in the 1D version of the system are the subject of the consideration in
Section VI. The paper is concluded by Section VII.

\section{The model}

In the usual mean-field approximation, the evolution of the spinor wave
functions of the spinor BEC, $\psi =(\psi _{+},\psi _{-})$, is governed by
the coupled Gross-Pitaevskii equations (GPEs), which include the SO coupling
of the Rashba type \cite{Rashba,Rashba2}:
\begin{eqnarray}
i\partial _{t}\psi _{\pm } &=&-{\frac{1}{2}}\nabla ^{2}\psi _{\pm }+\exp
\left( r^{2}\right) \cdot (|\psi _{\pm }|^{2}+\gamma |\psi _{\mp }|^{2})\psi
_{\pm }  \notag \\
&&\pm \lambda (\partial _{x}\mp i\partial _{y})\psi _{\mp }~.
\label{basicEq}
\end{eqnarray}%
The equations are written in the scaled form, with $\lambda $ being the
normalized strength of the SO-coupling, and $\gamma $ the relative strength
of the cross-interaction between the two components, with respect to the
self-repulsion. Strictly speaking, magnetic field, which imposes
the necessary Feshbach-resonance landscape, may differently
affect two hyperfine atomic states which compose the SO-coupled system.
In the present model, we neglect the difference, to admit analytical
solutions. Additional numerical results suggest that, if taken into regard,
the difference does not produce conspicuous changes in the results.

In physical units, the SO-coupling strength is characterized by ratio of the
respective length scale, $a_{\mathrm{SO}}$, to the confinement length, $a_{\perp }$,
which provides for the reduction of the 3D GPEs to the 2D form. In the real
experiment, this ratio is usually $a_{\mathrm{SO}}/a_{\perp }\sim 0.1$ \cite%
{YJLIN2011,Drummond}. The modulation profile of the repulsive nonlinearity
in Eq. (\ref{basicEq}) is adopted in the anti-Gaussian form, as it makes it
possible to find particular exact solutions for the self-trapped states, cf.
Ref. \cite{Borovkova20112}. Qualitatively, this steep profile does not lead
to results dramatically different from those produced by milder profiles
\cite{Borovkova2011,Borovkova20112}.

Stationary states with chemical potential $\mu $ are looked for as usual, $%
\psi _{\pm }(x,y,t)=\phi _{\pm }(x,y)e^{-i\mu t}$, where $\phi _{\pm }$ are
complex stationary wave functions. These states are characterized by the
total norm, proportional to the number of atoms in the binary BECs:
\begin{equation}
N=N_{+}+N_{-}=\int \int (|\psi _{+}|^{2}+|\psi _{-}|^{2})dxdy.  \label{Norm}
\end{equation}%
Further, the total energy of the soliton is
\begin{equation}
E=E_{\mathrm{K}}+E_{\mathrm{N}}+E_{\mathrm{SO}},
\end{equation}
where $E_{\mathrm{K}}$ is the kinetic term, while $E_{\mathrm{N}}$ and $E_{%
\mathrm{SO}}$ are energies of the nonlinear and SO interactions:
\begin{gather}
E_{\mathrm{K}}={\frac{1}{2}}\int d\mathbf{r}\left( |\nabla \psi
_{+}|^{2}+|\nabla \psi _{-}|^{2}\right) ,  \notag \\
E_{\mathrm{N}}={\frac{1}{2}}\int d\mathbf{r} e^{+r^2}\left(|\psi_{+}|^{4}+|%
\psi_{-}|^{4}+2\gamma|\psi_{+}|^{2}\|\psi_{-}|^{2}\right),  \notag \\
E_{\mathrm{SO}}=\lambda\int d\mathbf{r}\left[\psi _{+}^{\ast
}\left(\partial_{x}-i\partial_{y}\right)\psi _{-}-\psi _{-}^{\ast
}\left(\partial_{x}+i\partial_{y}\right)\psi _{+}\right].
\label{threeenergy}
\end{gather}

In the case of the SV solitons, $\psi _{+}$ and $\psi _{-}$ are defined as
the zero-vorticity and vortical components, respectively. Actually,
the SVs are the simplest axisymmetric fundamental modes existing in the system.
It is relevant to
define the relative share of the total norm carried by the vortex component:
\begin{equation}
F_{-}=N_{-}/N.  \label{N_}
\end{equation}

Stability of the solitons was investigated numerically by computing
eigenvalues for small perturbations, and the results were subsequently
verified by means of direct simulations of Eq. (\ref{basicEq}). To this end,
the perturbed solution was taken as
\begin{equation}
\psi _{\pm }=(\phi _{\pm }+u_{\pm }e^{-i\Lambda t}+v_{\pm }^{\ast
}e^{i\Lambda ^{\ast }t})e^{-i\mu t}.  \label{purterbation}
\end{equation}%
where $\ast $ stands for the complex conjugate. The substitution of this in
Eq. (\ref{basicEq}) and linearization leads to a system of four equations:

\begin{eqnarray}
(\mu +\Lambda )u_{+} &=&-{\frac{1}{2}}\nabla ^{2}u_{+}+\lambda (\partial
_{x}-i\partial _{y})u_{-}  \notag \\
&&+e^{r^{2}}\left( 2|\phi _{+}|^{2}+\gamma |\phi _{-}|^{2}\right) u_{+}
\notag \\
&&+e^{r^{2}}\left( \phi _{+}^{2}v_{+}+\gamma \phi _{+}\phi _{-}v_{-}+\gamma
\phi _{-}^{\ast }\phi _{+}u_{-}\right) ,  \notag \\
(\mu +\Lambda )u_{-} &=&-{\frac{1}{2}}\nabla ^{2}u_{-}-\lambda (\partial
_{x}+i\partial _{y})u_{+}  \notag \\
&&+e^{r^{2}}\left( 2|\phi _{-}|^{2}+\gamma |\phi _{+}|^{2}\right) u_{-}
\notag \\
&&+e^{r^{2}}\left( \phi _{-}^{2}v_{-}+\gamma \phi _{-}\phi _{+}v_{+}+\gamma
\phi _{+}^{\ast }\phi _{-}u_{+}\right) ,  \notag \\
(\mu -\Lambda )v_{+} &=&-{\frac{1}{2}}\nabla ^{2}v_{+}+\lambda (\partial
_{x}+i\partial _{y})v_{-}  \notag \\
&&+e^{r^{2}}\left( 2|\phi _{+}|^{2}+\gamma |\phi _{-}|^{2}\right) v_{+}
\notag \\
&&+e^{r^{2}}\left( \phi _{+}^{\ast 2}u_{+}+\gamma \phi _{+}^{\ast }\phi
_{-}^{\ast }u_{-}+\gamma \phi _{-}\phi _{+}^{\ast }v_{+}\right) ,  \notag \\
(\mu -\Lambda )v_{-} &=&-{\frac{1}{2}}\nabla ^{2}v_{-}-\lambda (\partial
_{x}-i\partial _{y})v_{+}  \notag \\
&&+e^{r^{2}}\left( 2|\phi _{-}|^{2}+\gamma |\phi _{+}|^{2}\right) v_{-}
\notag \\
&&+e^{r^{2}}\left( \phi _{-}^{\ast 2}u_{-}+\gamma \phi _{-}^{\ast }\phi
_{+}^{\ast }u_{+}+\gamma \phi _{+}\phi _{-}^{\ast }v_{+}\right) .  \notag \\
&&
\end{eqnarray}%
This system leads to the eigenvalue problem for the perturbation
eigenfrequency, $\Lambda =\Lambda _{\mathrm{r}}+i\Lambda _{\mathrm{i}}$,
written in the matrix form:
\begin{equation}
\left(
\begin{array}{cccc}
A_{11} & A_{12} & A_{13} & A_{14} \\
A_{21} & A_{22} & A_{23} & A_{24} \\
A_{31} & A_{32} & A_{33} & A_{34} \\
A_{41} & A_{42} & A_{43} & A_{44}%
\end{array}%
\right) \left(
\begin{array}{c}
u_{+} \\
u_{-} \\
v_{+} \\
v_{-}%
\end{array}%
\right) =\Lambda \left(
\begin{array}{c}
u_{+} \\
u_{-} \\
v_{+} \\
v_{-}%
\end{array}%
\right) ,  \label{eigenmatrix}
\end{equation}%
with matrix elements defined as
\begin{eqnarray}
&&A_{11}=-{\frac{1}{2}}\nabla ^{2}-\mu +e^{r^{2}}\left( 2|\phi
_{+}|^{2}+\gamma |\phi _{-}|^{2}\right) ,  \notag \\
&&A_{12}=\lambda (\partial _{x}-i\partial _{y})+e^{r^{2}}\gamma \phi
_{-}^{\ast }\phi _{+},  \notag \\
&&A_{13}=e^{+r^{2}}\phi _{+}^{2},\quad A_{14}=e^{+r^{2}}\gamma \phi _{+}\phi
_{-};  \notag \\
&&A_{21}=-\lambda (\partial _{x}+i\partial _{y})+e^{r^{2}}\gamma \phi
_{+}^{\ast }\phi _{-},  \notag \\
&&A_{22}=-{\frac{1}{2}}\nabla ^{2}-\mu +e^{r^{2}}\left( 2|\phi
_{-}|^{2}+\gamma |\phi _{+}|^{2}\right)  \notag \\
&&A_{23}=A_{14},\quad A_{24}=e^{r^{2}}\phi _{-}^{2},  \notag \\
&&A_{31}=-A_{13}^{\ast },\quad A_{32}=-A_{23}^{\ast },  \notag \\
&&A_{33}=-A_{11}^{\ast },\quad A_{34}=-A_{12}^{\ast }  \notag \\
&&A_{41}=-A_{14}^{\ast },\quad A_{42}=-A_{24}^{\ast },  \notag \\
&&A_{43}=-A_{21}^{\ast },\quad A_{44}=-A_{22}^{\ast }.  \label{matrixelement}
\end{eqnarray}%
The unperturbed solution $\phi _{\pm }$ is stable if all the eigenvalues $%
\Lambda $ are real.

\section{Semi-vortex (SV)\ solitons}

\subsection{The exact solution}

We adopt an ansatz for SV solutions to Eq. (\ref{basicEq}), with chemical
potential $\mu $, as
\begin{eqnarray}
&&\psi _{+}(x,y,t)=e^{-i\mu t}f_{1}(r^{2})  \notag \\
&&\psi _{-}(x,y,t)=e^{-i\mu t+i\theta }rf_{2}(r^{2}).  \label{twoansatz}
\end{eqnarray}%
where $(r,\theta )$ are the polar coordinates, and real functions $%
f_{1,2}(r^{2})$ obey the following equations:
\begin{eqnarray}
\mu f_{1} &+&2\left( r^{2}f_{1}^{~\prime \prime }+f_{1}^{~\prime }\right)
-\exp \left( r^{2}\right) \cdot (f_{1}^{~2}+\gamma r^{2}f_{2}^{~2})f_{1}
\notag \\
&&-2\lambda (r^{2}f_{2}^{~\prime }+f_{2})=0  \notag \\
\mu f_{2} &+&2\left( r^{2}f_{2}^{~\prime \prime }+2f_{2}^{~\prime }\right)
-\exp \left( r^{2}\right) \cdot (r^{2}f_{2}^{~2}+\gamma f_{1}^{~2})f_{2}
\notag \\
&&+2\lambda f_{1}^{~\prime }=0,  \label{polarGP}
\end{eqnarray}%
where
\begin{equation}
f_{1,2}^{~\prime }\equiv {\frac{d}{d(r^{2})}}f_{1,2}\quad f_{1,2}^{~\prime
\prime }\equiv {\frac{d^{2}}{d(r^{2})^{2}}}f_{1,2}.
\end{equation}%
Further, assuming
\begin{equation}
f_{1,2}=A_{\pm }\exp \left( -r^{2}/2\right) ,  \label{guessfunction}
\end{equation}%
and substituting this in Eq. (\ref{polarGP}), it is easy to see that Eqs. (%
\ref{twoansatz}) and (\ref{guessfunction}) indeed produce an \emph{exact
solution}, provided that the constants are expressed in terms of $\gamma $
as follows: 
\begin{eqnarray}
&&\lambda ^{2}={\frac{(1-\gamma )(2-\gamma )}{4}},  \label{lambda} \\
&&\mu ={\frac{(2-\gamma )}{2(1-\gamma )}}+\gamma ,  \label{finalmu} \\
&&A_{+}=\left[ {\frac{2-\gamma }{2(1-\gamma )}}\right] ^{1/2}.
\label{finalA1} \\
&&A_{-}^{2}=1/2.  \label{A2}
\end{eqnarray}%
In particular, Eq. (\ref{lambda}) implies that the present exact solution is
a particular one (rather than being generic), because it exists only if the
SO-coupling strength is adjusted to the value fixed by Eq. (\ref{lambda}).
With regard to these results, the norm of each component of the soliton is
\begin{gather}
N_{+}={\frac{\pi (2-\gamma )}{2(1-\gamma )},}  \label{N1} \\
N_{-}={\frac{\pi }{2},}  \label{N2} \\
N=N_{+}+N_{-}={\frac{\pi }{2}}\left( {\frac{2-\gamma }{1-\gamma }}+1\right) ,
\label{totalnorm}
\end{gather}%
and the share of the total norm in the vortical component [see Eq. (\ref{N_}%
)] is
\begin{equation}
F_{-}={\frac{1-\gamma }{3-2\gamma }}.  \label{FN}
\end{equation}

Equations (\ref{lambda})-(\ref{totalnorm}) suggest that the exact solution
is solely controlled by $\gamma $, and moreover, Eq. (\ref{lambda})
demonstrates that the exact solution does not exists for $1\leq \gamma \leq 2
$. According to Eqs. (\ref{N1}) and (\ref{N2}), for $0\leq \gamma <1$ (the
cross-repulsion is weaker than the self-repulsion), one has $N_{+}>N_{-}$
(the fundamental component is the larger one); however, for $\gamma >2$, the
situation is opposite, featuring a larger vortex component: $N_{+}<N_{-}$.
\begin{figure}[tbh]
{\includegraphics[width=0.6\columnwidth]{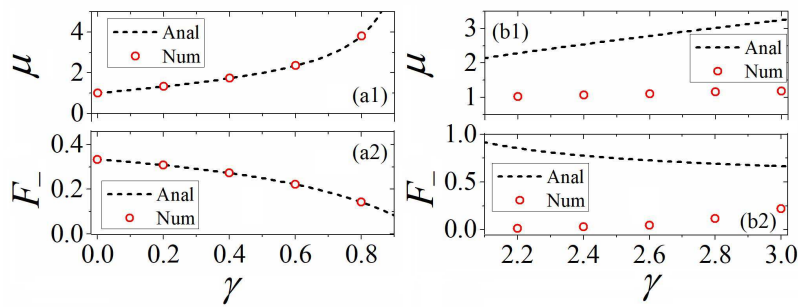}}
\caption{(a1,a2) Comparison between the analytical and numerical results
(the black dashed curve and red circles, respectively) for the soliton's
chemical potential, $\protect\mu $, and the norm share in the vortical
component, $F_{-}$, in the region of $0\leq \protect\gamma <1$. (b1,b2) The
same for $\protect\gamma >2$. Parameters $\protect\lambda $ (the strength of
the SO coupling) and $N$ (the total norm) are here taken as per by Eqs. (%
\protect\ref{lambda}) and (\protect\ref{totalnorm}), respectively. The
dashed curves for $\protect\mu (\protect\gamma )$ and $F_{-}(\protect\gamma )
$ are plotted, severally, as per Eqs. (\protect\ref{finalmu}) and (\protect
\ref{FN}). }
\label{NumAna}
\end{figure}
\begin{figure}[tbh]
{\includegraphics[width=0.9\columnwidth]{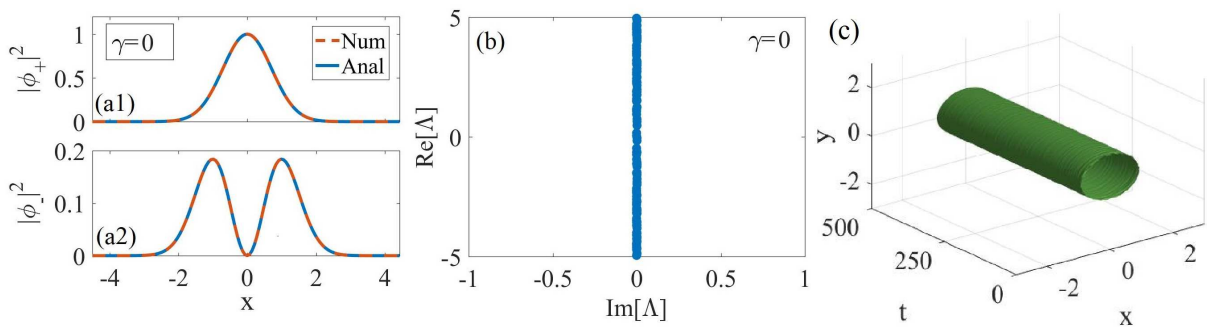}}
\caption{(a1,a2) The comparison between the analytical and numerically found
(blue solid and red dashed curves) stationary profiles of the soliton's wave
function, shown in cross-section $y=0$, at $\protect\gamma =0$. (b) Spectra
of stability eigenvalues $\Lambda $ for the exact solution from panels
(a1,a2). (c) Direct simulations of the evolution of this soliton (with
2\% random noise added to the initial conditions), shown by the contour plot of the density profile,
$|\protect\psi _{+}(\mathbf{r},t)|^{2}$. In this figure, the analytical wave
function is taken as per Eqs. (\protect\ref{twoansatz}), (\protect\ref%
{guessfunction}), (\protect\ref{finalA1}) and (\protect\ref{A2}), while $%
\protect\lambda $ and $N$ are given by Eqs. (\protect\ref{lambda}) and (%
\protect\ref{totalnorm}). }
\label{stableexact}
\end{figure}

\begin{figure}[tbh]
{\includegraphics[width=0.9\columnwidth]{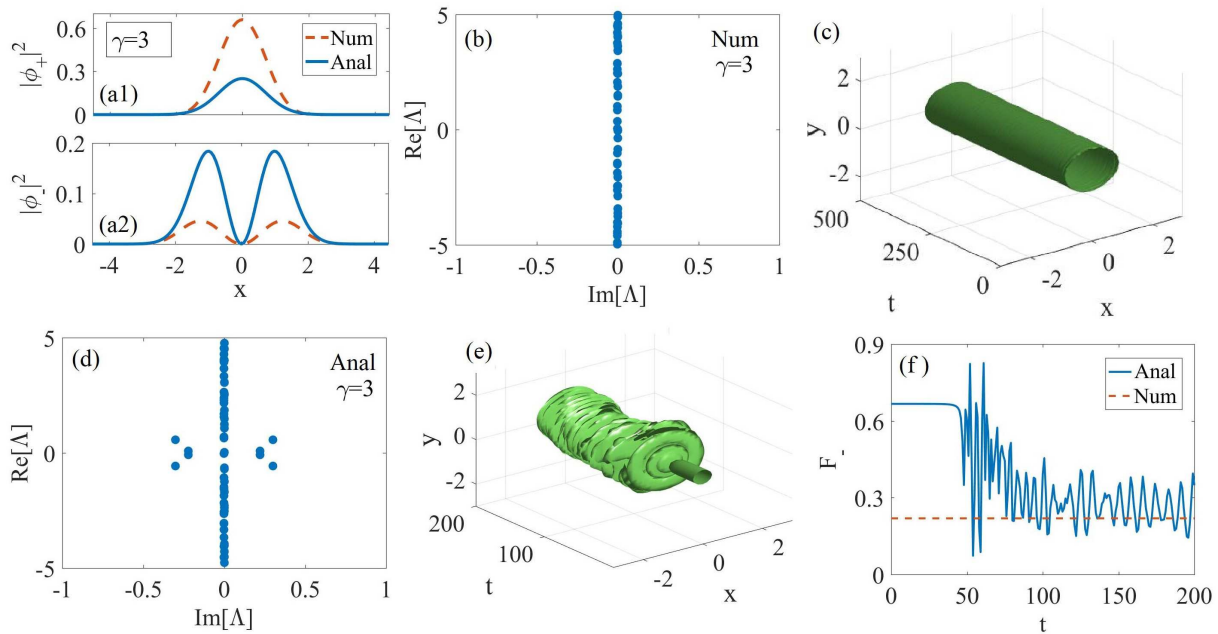}}
\caption{(a1,a2) Comparison between the analytical solution and numerically
found ground state results (the blue solid and red dashed curves,
respectively), shown in cross-section $y=0$, at $\protect\gamma =3$. (b)
Spectra of the stability eigenvalues, $\Lambda$, for the numerical solution
(the one represented by the red dashed curves in panels (a1,a2). (c) Direct
simulations to the evolution of the numerical solution (with 2\% random noise
added to the initial conditions), shown by the contour plot of the density profile,
$|\protect\psi _{+}(\mathbf{r},t)|^{2}$. (d) Spectra of the stability
eigenvalues $\Lambda $ for the analytical solution (represented by the blue
solid curve in panels (a1,a2). (e) Direct simulations of the perturbed
evolution of the analytical solution, shown by the contour plot of the
density profile, $|\protect\psi _{+}(\mathbf{r},t)|^{2}$. (f) $F_{-}$,
defined as per Eq. (\protect\ref{N_}), as observed in the direct simulations
initiated by the exact solution and the numerical one (the blue solid and
red dashed curves, respectively). The analytical solution dealt with in this
figure is taken as per Eqs. (\protect\ref{twoansatz}), (\protect\ref%
{guessfunction}), (\protect\ref{finalA1}) and (\protect\ref{A2}), while
parameters $\protect\lambda $ and $N$ are defined by Eqs. (\protect\ref%
{lambda}) and (\protect\ref{totalnorm}).}
\label{unstableexact}
\end{figure}

For comparison with the analytical results, we used numerically found
solutions for the ground state of the present system, which were generated,
for the same values of parameters, by means of the imaginary-time method.
The comparison for the soliton's chemical potential and the norm share of
the vortical component, $F_{-}$ [see Eq. (\ref{N_})], in the regions of $%
0\leq \gamma <1$ and $\gamma >2$ is displayed in Fig. \ref{NumAna}. Figures %
\ref{NumAna}(a1,a2) show that the analytical solution exactly produces the
ground state for $0\leq \gamma <1$. The comparison of the respective
wave-function profiles, as produced by the numerical and exact solutions in
this case, is presented in Figs. \ref{stableexact}(a1,a2). Further, Figs. %
\ref{stableexact}(b) and (c) corroborate the stability of the exact
solution, which should be expected from the ground state.

However, in the region of $\gamma >2$, Figs. \ref{NumAna}(b1,b2) and \ref%
{unstableexact}(a1,a2) show that the analytical solution is different from
the ground states. In particular, Fig. \ref{NumAna}(b2) shows that the
numerically found ground state always has $F_{-}<0.5$ (i.e., the
zero-vorticity components remains the dominant one), while the exact
solution has $F_{-}>0.5$ at $\gamma \geq 2$. Thus, the exact solution
represents an excited state of the SV, which is unstable, as seen in Figs. %
\ref{unstableexact}(d-f), on the contrary to the stability of the
numerically found ground state, corroborated by Figs. \ref{unstableexact}%
(b,c). Lastly, the evolution of the norm share in the vortical component, $%
F_{-}(t)$, displayed in Fig. \ref{unstableexact}(f), suggests that the
evolution of the unstable exact SV at $\gamma >2$ tends to transform it
towards the stable ground state. We note that full soliton family in the
system depends not only on $\gamma$, but also on other control parameters,
{\it viz}., $N$ and $\lambda$. Thus, the exact
solution produces only specific subfamilies of the solitons, possible
links between which are not available in the analytical form. In the following
subsection, we produce numerical results for a broader soliton family,
into which the analytical branches may be embedded.

\subsection{Full numerical results}

In the numerical form, soliton solutions were obtained by means of the
imaginary-time method. To this end, the same input was
used as in Ref. \cite{SVS1}, which agrees with the general SV ansatz (\ref%
{twoansatz}):%
\begin{equation}
\phi _{+}^{(0)}\left( r,\theta \right) =A_{+}\exp (-\alpha _{+}r^{2}),~\phi
_{-}^{(0)}\left( r,\theta \right) =A_{-}r\exp (i\theta -\alpha _{-}r^{2}),
\label{SVguess}
\end{equation}%
with real constants $A_{\pm }$ and $\alpha _{\pm }$. The control parameters
are $\allowbreak N$, $\gamma $, and $\lambda $.

\begin{figure}[tbh]
{\includegraphics[width=0.6\columnwidth]{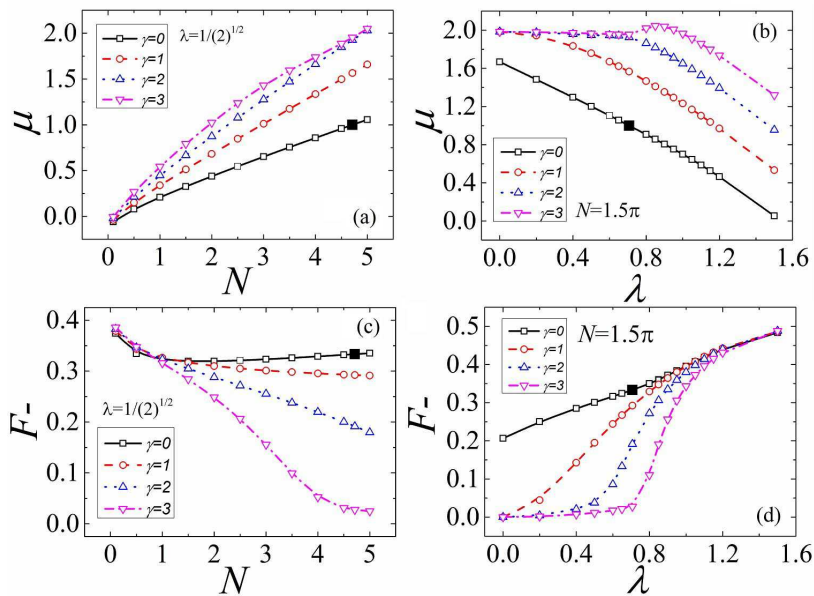}}
\caption{{(a,b)} The chemical potential of the numerically found family of
semi-vortices, $\protect\mu $, versus the total\ norm, $N$, and the
SO-coupling strength, $\protect\lambda $, for different values of the
cross/self interaction ratio, $\protect\gamma =0$, $1$, $2$, $3$,
respectively. (c,d) The norm share in the vortical component, $F_{-}$ [see
Eq. (\protect\ref{N_})], versus $N$ and $\protect\lambda $ for $\protect%
\gamma =0$, $1$, $2$, and $3$, respectively. In panels (a,c) and (b,d), $%
\protect\lambda =1/\protect\sqrt{2}$ and $N=1.5\protect\pi$ is fixed,
severally. The black solid squares in the panels represent the exact
solution, given by Eqs. (\protect\ref{twoansatz}) and (\protect\ref{lambda}%
)-(\protect\ref{FN}), with $\protect\gamma=0$.}
\label{SVmuFN}
\end{figure}

Figure \ref{SVmuFN}(a) displays the chemical potential of the SV family
versus $N$ for different values of $\gamma $. The $\mu(N)$
curves feature a positive slope, with $d\mu /dN>0$, i.e., they satisfy
the {\it anti-Vakhitov-Kolokolov} criterion, which is conjectured to provide a
necessary stability condition for bright
solitons in self-repulsive nonlinear media \cite{Fukuoka}. SVs with larger
values of $\gamma $ have larger $\mu $, which also is a natural corollary of
the repulsive sign of the nonlinearity. The negative slope, $d\mu /d\lambda
<0$, in Fig. \ref{SVmuFN}(b), for different values of $\gamma $, implies
that the SO-coupling energy is negative, partly compensating the repulsive
nonlinear interactions. The decay of $F_{-}(N)$ with the increase of $N$,
observed in Fig. \ref{SVmuFN}(c), shows the same trend as reported for
stable SVs in the self-attractive system \cite{SVS1}, i.e., concentration of
the norm in the zero-vorticity component. Note that the $F_{-}(N)$ curves of
$F_{-}(N)$ with different values of $\gamma $ are degenerate at small $N$,
splitting with the increase of $N$, the vortical-component's share, $F_{-}$,
decaying faster under the action of stronger cross-repulsion (for larger $%
\gamma $). On the other hand, Fig. \ref{SVmuFN}(d) demonstrates the growth
of $F_{-}$ with the increase of $\lambda $, which is natural too, as
stronger SO coupling generates more vorticity, i.e., it causes transfer of
the norm to the vortical component. In particular, $F_{-}$ becomes nearly
independent of $\gamma $ at $\lambda >1$, which means that the SO
interaction between the components dominates over the nonlinear repulsion
between them in this case. In Fig. \ref{SVmuFN}, the exact analytical
solutions, included in the families of numerically generated ones, are
marked by back solid squares.

Lastly, we stress that the SV\ family, produced by the imaginary-time
integration, is found to be completely stable (in terms of stability
eigenvalues and direct simulations alike) at all values of the parameters,
including all values of $\gamma $. The latter conclusion is essential,
because, in the 2D SO-coupled system with attractive interactions, the SVs
are stable only at $\gamma \leq 1$ \cite{SVS1}.


\section{Fundamental mixed-mode (MM) solitons}

\begin{figure}[tbh]
{\includegraphics[width=0.65\columnwidth]{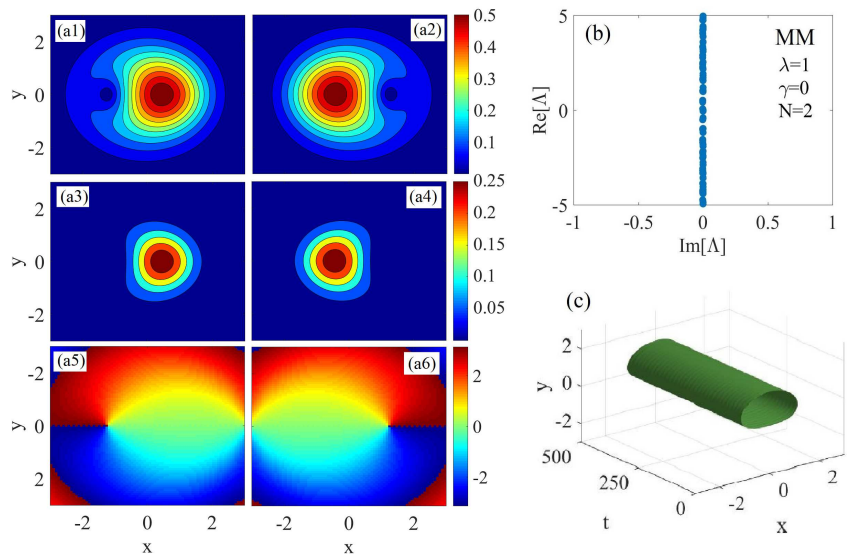}}
\caption{(a) A typical example of a stable MM (mixed-mode) soliton, shown by
$|\protect\phi _{\pm}(\mathbf{r})|$ (a1,a2), $|\protect\phi _{\pm}(\mathbf{r}%
)|^{2}$(a3,a4) as well as their phase diagrams (a5,a6), respectively, with $%
(N,\protect\gamma ,\protect\lambda )=(2,0,1)$. (b) The spectrum of stability
eigenvalues for this state. (d) Direct simulation of its evolution with
2\% noise, shown by means of the density plot, $|\protect\phi _{+}(\mathbf{r}%
,t)|^{2}$.}
\label{ExpMM}
\end{figure}

Following Ref. \cite{SVS1}, MM states can be initiated by the input which
includes terms with vorticities $(0,-1)$ and $(0,+1)$ in the two components,
\begin{equation}
\phi _{\pm }^{(0)}(r,\theta )=A_{1}\exp (-\alpha _{1}r^{2})\mp A_{2}r\exp
(-\alpha _{2}r^{2}\mp i\theta ),  \label{MMguess}
\end{equation}%
where $A_{1,2}$ and $\alpha _{1,2}$ are real constants, cf. Eq. (\ref%
{SVguess}). The initial approximation for the MM states are
built as equal-weight superpositions of SVs with topological content $(0,-1)$
and $(0,+1)$ in the two components. In a certain sense, the SVs and MMs are
similar, respectively, to immiscible and miscible states in binary
superfluids. Unlike the SVs, MMs cannot be represented by an exact ansatz,
but numerical and variational results clearly confirm their existence.
A typical example of a stable MM with $(N,\gamma ,\lambda )=(2,0,1)$,
generated by the imaginary-time integration, is shown in Fig. \ref{ExpMM}.

The numerical analysis demonstrates that, similar to what is reported above
for the SVs, the family of the MM states generated by the imaginary-time
simulations from input (\ref{MMguess}) is completely stable at all values of
parameters, including, in particular, all values of $\gamma $. This
conclusion is essential because, in the 2D\ SO-coupled system with
attractive interactions, MM states are stable solely at $\gamma \geq 1$ \cite%
{SVS1}, i.e., the MMs are stable, in the care of the attractive
nonlinearity, where the SVs are not, and vice versa, coexisting as stable
states solely in the case of the Manakov's nonlinearity \cite{Manakov},
i.e., at $\gamma =1$. The stability switch between the SVs and MMs was
explained in Ref. \cite{SVS1} by the fact that, for equal values of $N$, the
SV has a smaller energy, i.e., it realizes the system's ground state, at $%
\gamma <1$, while at $\gamma >1$ the MM provides for the smaller energy,
i.e., the ground state. The SV and MM states are degenerate, having equal
energies, at $\gamma =1$, when both are stable.

\begin{figure}[tbh]
{\includegraphics[width=0.6\columnwidth]{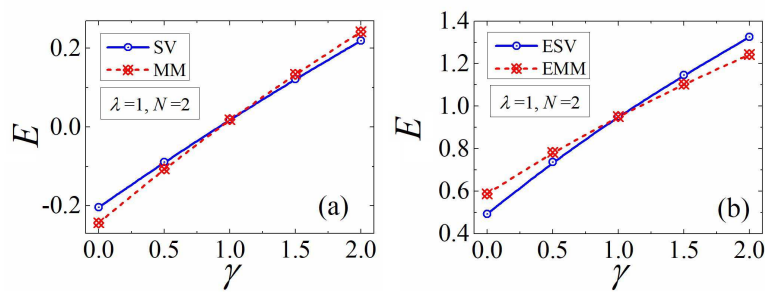}}
\caption{(a) The total energy of the SV (semi-vortex) and MM (mixed-mode)
states (blue solid and dashed red curves, respectively) versus $\protect%
\gamma $ for a fixed norm, $N=2$, and a fixed strength of the SO-coupling, $%
\protect\lambda =1$. (b) The Energy of ESV (excited-state of SV) with $%
S_{+}=1$ and EMM (Excited state of MM) with $S_{1}=1$ versus $\protect\gamma
$ for fixed $N=2$ and $\protect\lambda =1$. }
\label{muMM}
\end{figure}

In the present system, as said above, both the SVs and MMs are completely
stable at all values of $\gamma $. This conclusion is natural because the
system with repulsive interactions has a definite trend to be more stable
than its attractive counterpart. Nevertheless, it makes sense to identify
the ground state of the present system too, comparing values of energy of
the coexisting SV and MM\ solitons with equal values of the norm. The
comparison is shown in Fig. \ref{muMM}(a), which clearly demonstrates that
the SV\ and MM realize the system's ground state at $\gamma >1$ and $\gamma
<1$, respectively, which is precisely opposite to the situation in the 2D
SO-coupled system with the attractive interactions \cite{SVS1}. The same
conclusion was obtained for all other values of parameters, i.e., fixed $N$
and $\lambda $.

\section{Excited states}

\begin{figure}[tbh]
{\includegraphics[width=0.95\columnwidth]{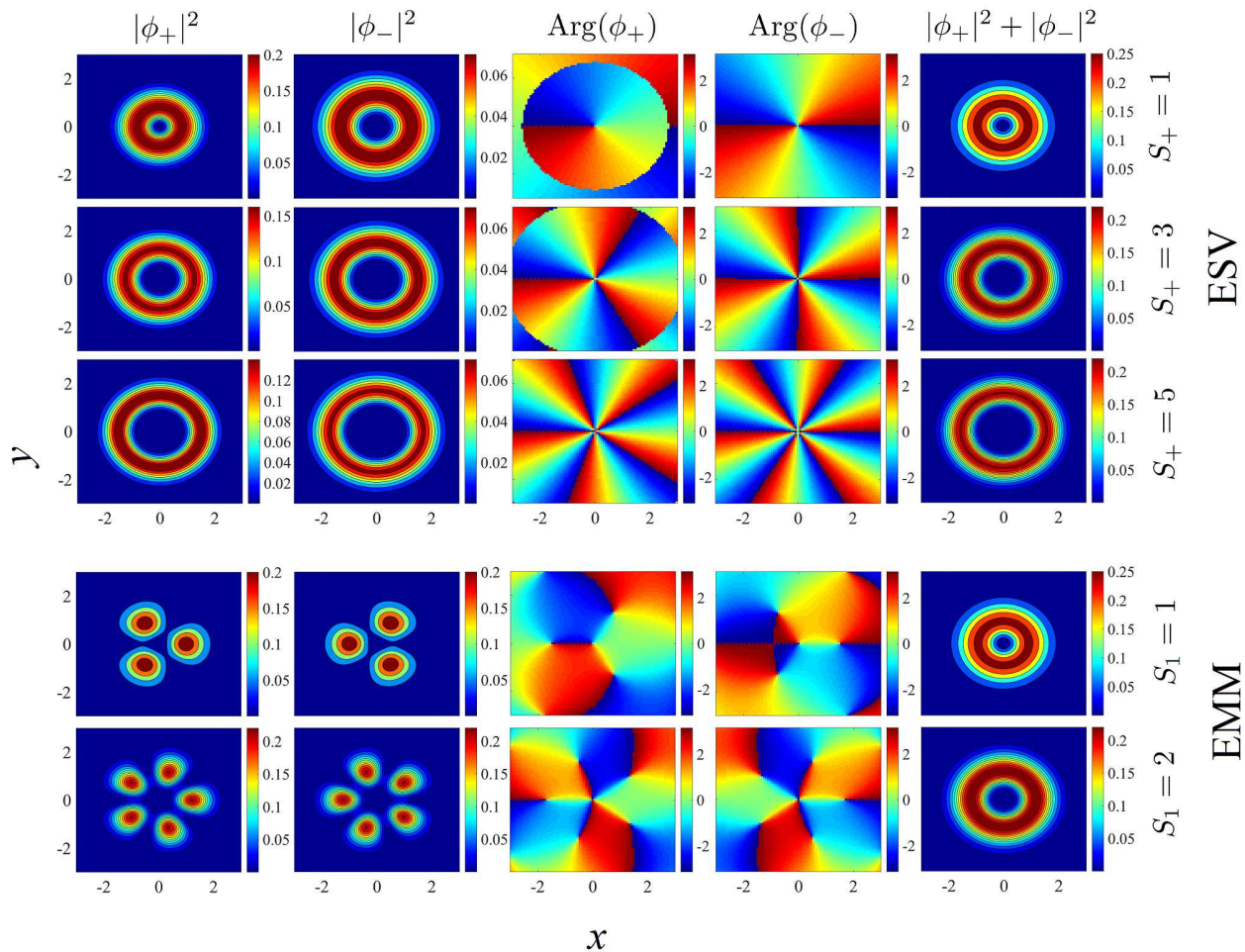}}
\caption{Examples of stable excited states of SVs (``ESVs") with $S_{+}=1$, $%
3$, and $5$ (the first, second, and third rows, respectively), and of MMs
(``EMMs") with $S_{1}=1$ and $2$ (the fourth and fifth rows, respectively).
The first and second columns display density patterns of components $\protect%
\phi _{\pm }$, while the third and fourth columns display phase structures
of $\protect\phi _{\pm }$. The fifth column is the total-density pattern, $n(%
\mathbf{r})=|\protect\phi _{+}|^{2}+|\protect\phi _{-}|^{2}$. Parameters are
$N=2$, $\protect\gamma =1$ and $\protect\lambda =1$ for all the panels. }
\label{ESexmaple}
\end{figure}
\begin{figure}[tbh]
{\includegraphics[width=0.6\columnwidth]{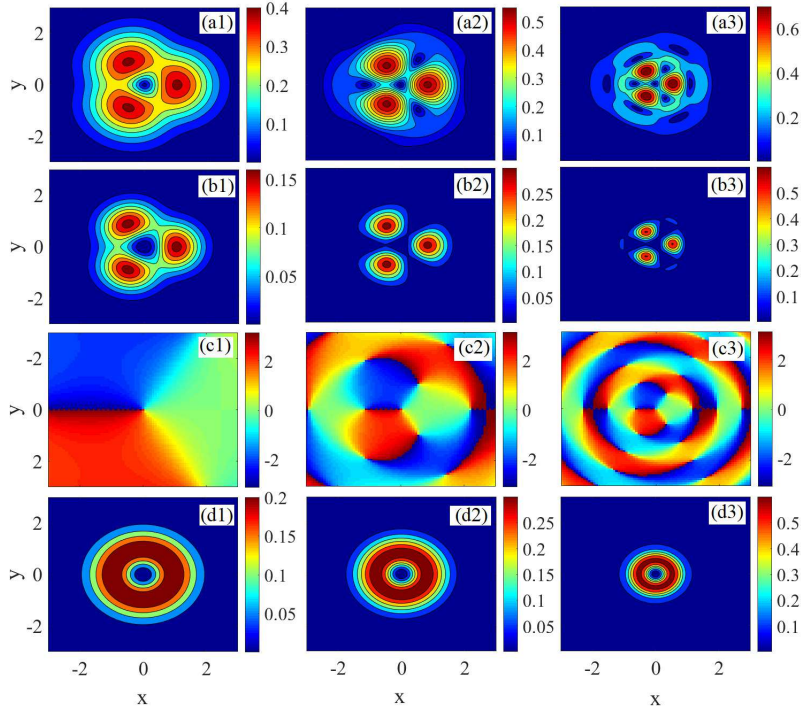}}
\caption{(a1-a3)$|\protect\phi_{+}|$ with $\protect\lambda=0.2$ (a1), 2 (a2)
and 4 (a3). (b1-b3) $|\protect\phi_{+}|^{2}$ with $\protect\lambda=0.2$
(b1), 2 (b2) and 4 (b3). (c1-c3)The phase diagram of $\protect\phi_{+}$ with
$\protect\lambda=0.2$ (c1), 2 (c2) and 4 (c3). (d1-d3) Total density pattern
$n(\mathbf{r})=|\protect\phi_{+}(\mathbf{r})|^{2}+|\protect\phi_{-}(\mathbf{r%
})|^{2}$ with $\protect\lambda=0.2$ (d1), 2 (d2) and 4 (d3). Total norm and $%
\protect\gamma$ of these examples are fixed by $(N,\protect\gamma)=(2,1)$.
Solitons in these panels are all stable.}
\label{ESexmaplevslam}
\end{figure}
\begin{figure}[tbh]
{\includegraphics[width=0.9\columnwidth]{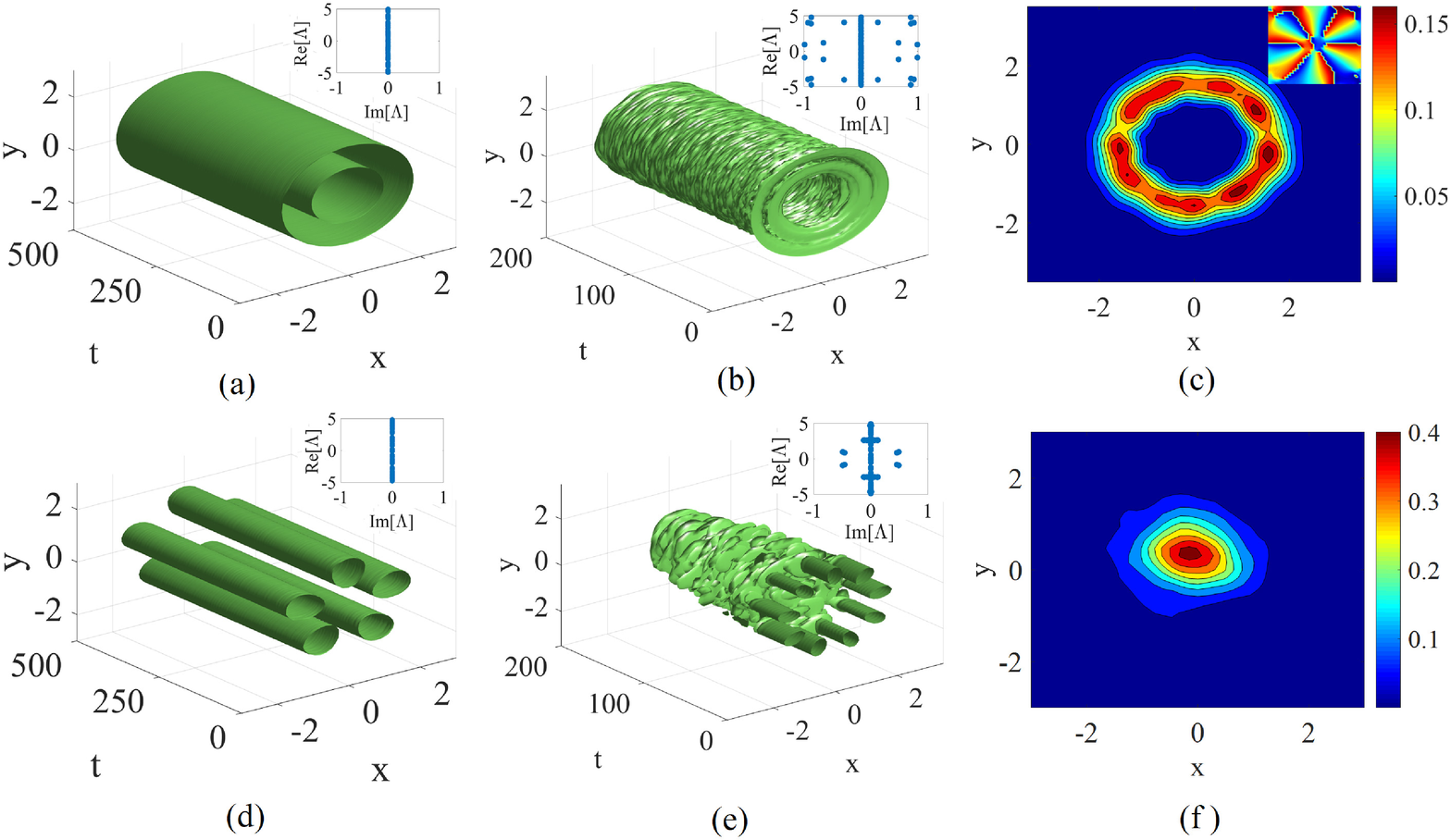}}
\caption{Example of the real-time evolution of stable and unstable excited
states of SVs and MMs. (a) A stable SV's excited state with $(N,\protect%
\gamma ,\protect\lambda ,S_{+})=(2,1,1,4)$. (b) An unstable SV state with $%
(N,\protect\gamma ,\protect\lambda ,S_{+})=(2,1,1,6)$. (c) The final density
pattern of $|\protect\psi _{+}|^{2}$, corresponding to panel (b), the inset
showing the respective phase pattern. (d) A stable excited state of the MM
with $(N,\protect\gamma ,\protect\lambda ,S_{1})=(2,1,1,2)$. (e) An unstable
MM state with $(N,\protect\gamma ,\protect\lambda ,S_{1})=(2,1,1,4)$. (f)
The final density pattern of $|\protect\psi _{+}|^{2}$, corresponding to
panel (e). Spectra of stability eigenvalues for these modes are shown as
insets in panels (a), (b), (d) and (e).}
\label{PropagationEs}
\end{figure}

Excited states of SVs and MMs are produced by adding extra vorticity to both
components of these states. In particular, as suggested by Ref. \cite{SVS1},
excited states of the SVs correspond to the following ansatz compatible with
Eq. (\ref{basicEq}), cf. Eq. (\ref{twoansatz}):
\begin{eqnarray}
&&\psi _{+}(x,y,t)=e^{-i\mu t+in\theta }r^{n}f_{1}(r^{2})  \notag \\
&&\psi _{-}(x,y,t)=e^{-i\mu t+i(n+1)\theta }r^{n+1}f_{2}(r^{2}),
\label{ansatz}
\end{eqnarray}%
with $n=1,2,...$ . However, in the previously studied systems, all the
excited states, unlike the fundamental MMs and SVs, were found to be
completely unstable in SO-coupled systems with homogeneous contact
nonlinearities \cite{SVS1,QD}. Very recently, stable excited states of SVs
and MMs were predicted in the model of the 2D dipolar BEC with long-range
interactions between field-induced dipoles whose magnitude grows from the
center to periphery \cite{ESsoc}. The latter finding suggests to explore the
existence and stability of excited in the present system with the contact
repulsive interactions.

We have performed the analysis, running imaginary-time simulations initiated
by the following inputs for the excited states of SVs:%
\begin{equation}
\phi _{\pm }^{(0)}=A_{\pm }r^{|S_{\pm }|}\exp (-\alpha _{\pm }r^{2}+iS_{\pm
}\theta ),  \label{ESSV}
\end{equation}%
cf. the general exact ansatz (\ref{ansatz}), and for the excited states of
MMs:
\begin{gather}
\phi _{\pm }^{(0)}=A_{1}r^{|S_{1}|}\exp (-\alpha _{1}r^{2}\pm iS_{1}\theta )
\notag \\
\mp A_{2}r^{|S_{2}|}\exp (-\alpha _{2}r^{2}\mp iS_{2}\theta ),  \label{ESMM}
\end{gather}%
where $A_{\pm }$, $A_{1,2}$, $\alpha _{\pm }$ and $\alpha _{1,2}$ are real
constants, $S_{+,1}=0,\pm 1,\pm 2,\cdots $ are integers, and $%
S_{-,2}=S_{+,1}+1$. In Eq. (\ref{ESSV}), fundamental SVs are produced by $%
(S_{+},S_{-})=(-1,0)$ or $(0,1)$, cf. Eq. (\ref{SVguess}), and their excited
states correspond to $(S_{+},S_{-})=(n,n+1)$, with $n\neq -1$ or $0$.
Similarly, Eq. (\ref{ESMM}) produces fundamental MMs for $S_{1}=-1$ and $0$,
and their excited states correspond to other integer values of $S_{1}$.
Below, we fix $\lambda =1$ and consider $S_{+,1}~\geq 1$, for $N\leq 5$ and $%
\gamma \leq 2$.

Numerical results demonstrate that embedded states of SVs may be stable up
to $S_{+}=5$, while MMs in the excited states are stable only up to $S_{1}=2$%
. Examples of the 2D excited states are displayed in Fig. \ref{ESexmaple}.
In the case of the SVs, they feature a standard ring structure in each
components (see top three rows in Fig. \ref{ESexmaple}), with the
topological charges identical to the values of $S_{\pm }$ in input (\ref%
{ESSV}). In the case of MMs, each component of the excited state is built as
annular necklace structures, the number of fragments in the necklaces being
exactly equal to $S_{1}+S_{2}=2S_{1}+1$, see two bottom rows in Fig. \ref%
{ESexmaple}. The total angular momentum of the MM states is $L=\left(
L_{+}+L_{-}\right) /2\equiv 0$. The total density patterns, i.e. $|\phi _{+}(%
\mathbf{r})|^{2}+|\phi _{-}(\mathbf{r})|^{2}$, of the excited states of both
SVs and MMs exhibit perfect ring patterns. The dependence of the
necklace pattern of the excited MM states on the SO strength, $\lambda $,
is illustrated by Fig. \ref{ESexmaplevslam}, which displays excited MMs with $%
(N,\gamma )=(2,1)$ and different values of $\lambda $. The figure shows
that the main density pattern of the excited MM states shrinks with the
increase of $\lambda $, and higher-order lobes emerge from pattern's core.
Recently, somewhat similar 2D necklace patterns
were realized by the SO-coupled BEC in an annular trapping potential \cite%
{white2017}, where they are ground states, featuring a dependence on $%
\lambda $ different from that produced by the excited MM states in the
current work.

Typical example of the evolution of stable and unstable excited states are
displayed in Fig. \ref{PropagationEs}. As mentioned above, excited states of
SVs are stable for $S_{+}\leq 5$. At $S_{+}>5$, the instability perturbs the
SV pattern, but does not destroy its vortex structure. Excited states of MMs
are stable only for $S_{1}=1$ and $2$. Unstable MM\ excited states tend to
transform into fundamental MMs.

The different instability evolution of the SV and MM excited states is
explained by the presence of the angular momentum in SVs. As a result,
unstable excited states of SVs keep its initial value, which prevents their
transformation into their stable counterparts with smaller values of the
angular momentum [see an example in Figs. \ref{PropagationEs}(b,c), where
the output phase pattern demonstrate that the vorticity of $\psi _{+}$%
remains $6$]. On the other hand, the zero total angular momentum of unstable
MM excited states allows them to simplify themselves into the fundamental
MM, see an example in Fig. \ref{PropagationEs}(e,f).

Finally, families of the excited states of SVs and MMs are characterized by
the corresponding $E(\gamma)$ dependence, which is displayed in Fig. \ref%
{muMM}(b) with fixed values of $S_{+}$ and $S_{1}$ for $N=2$. This panel
shows that the first excited states of SVs and MMs are degenerate, having
equal energies, at $\gamma=1$. The SV excited states have a lower energy
than their counterparts of the MM type at $\gamma<1$, and vice versa at $%
\gamma>1$.

\section{Soliton in the 1D model}

The 1D reduction of the 2D system (\ref{basicEq}) amounts to the following
system of coupled GPEs:
\begin{equation}
i\partial _{t}\psi _{\pm }=-{\frac{1}{2}}\partial _{xx}^{2}\psi _{\pm }+\exp
\left(+x^{2}\right) \cdot (|\psi _{\pm }|^{2}+\gamma |\psi _{\mp }|^{2})\psi
_{\pm }\pm \lambda \partial _{x}\psi _{\mp }~.  \label{1D}
\end{equation}%
Stationary solutions with chemical potential are looked for in the usual
form, $\psi _{\pm }(x,t)=e^{-i\mu t}u_{\pm }(x)$, where real functions $%
u_{\pm }(x)$ (unlike complex stationary wave function in the 2D case) solve
equations%
\begin{eqnarray}
\mu u_{+}+{\frac{1}{2}}u_{+}^{\prime \prime }-e^{x^{2}}(u_{+}^{2}+\gamma
u_{-}^{2})u_{+}-\lambda u_{-}^{\prime } &=&0,  \label{u+} \\
\mu u_{-}+{\frac{1}{2}}u_{-}^{\prime \prime }-e^{x^{2}}(u_{-}^{2}+\gamma
u_{+}^{2})u_{-}+\lambda u_{+}^{\prime } &=&0,  \label{u-}
\end{eqnarray}%
where the prime stands for $d/dx$. A particular exact solution to Eqs. (\ref%
{u+}) and (\ref{u-}) with one even and one odd components, which may be
considered as a 1D counterpart of the 2D SV states, is looked for in the
form of%
\begin{equation}
u_{+}=A_{+}\exp \left( -x^{2}/2\right) ,~u_{-}=A_{-}x\exp \left(
-x^{2}/2\right) .  \label{uu}
\end{equation}%
Equation (\ref{uu}) implies that component $u_{-}$ has a dipole structure,
therefore, modes of this type may be named semi-dipole solitons.
Substituting this in Eqs. (\ref{u+}) and (\ref{u-}), we obtain the following
relations between the constants:
\begin{eqnarray}
&&\mu A_{+}-{\frac{1}{2}}A_{+}-A_{+}^{3}-\lambda A_{-}=0,  \notag \\
&&{\frac{A_{+}}{2}}-\gamma A_{-}^{2}A_{+}+\lambda A_{-}=0,  \notag \\
&&\mu A_{-}-{\frac{3}{2}}A_{-}-\gamma A_{+}^{2}A_{-}-\lambda A_{+}=0,  \notag\\
&&{\frac{1}{2}}A_{-}-A_{-}^{3}=0.  \label{1D4equation}
\end{eqnarray}%
Solving Eq. (\ref{1D4equation}), one may obtain
\begin{eqnarray}
&&\lambda ^{2}={\frac{(\gamma -1)(\gamma -3)}{8}},  \label{lambda1D} \\
&&\mu ={\frac{(\gamma +1)(2\gamma -3)}{4(\gamma -1)}},  \label{mu1D} \\
&&A_{+}^{2}={\frac{(\gamma -3)}{4(\gamma -1)}},  \label{APoneD} \\
&&A_{-}^{2}={\frac{1}{2}}.  \label{ANoneD}
\end{eqnarray}%
Equation (\ref{lambda1D}) indicates that the exact solution exist in the
cases of $\gamma <1$ and $\gamma >3$. The norms of the two components of the
exact soliton are%
\begin{eqnarray}
&&N_{+}={\frac{\sqrt{\pi }}{4}}\left( {\frac{\gamma -3}{\gamma -1}}\right) ,\label{N+1D} \\
&&N_{-}={\frac{\sqrt{\pi }}{4}}.  \label{N-1D} \\
&&N=N_{+}+N_{-}={\frac{\sqrt{\pi }}{2}}\left( {\frac{\gamma -2}{\gamma -1}}%
\right) .  \label{totalnorm1D}
\end{eqnarray}%
Therefore, the 1D version of the analytical result for the relative share of
the total norm which is kept in the odd component is
\begin{equation}
F_{-}={\frac{\gamma -1}{2(\gamma -2)}},  \label{FN1D}
\end{equation}%
cf. Eq. (\ref{N_}). Similar to the exact 2D solution, parameters of the 1D
solution, $\lambda $ and $N$, are solely defined by $\gamma $ [cf. Eq. (\ref%
{lambda}) and (\ref{totalnorm})]. Moreover, according to Eqs. (\ref{FN1D}),
if $0\leq \gamma <1$, then $F_{-}<0.5$ (the fundamental component is the
larger one); however, if $\gamma >3$, then $F_{-}>0.5$ (the vortex component
is larger).

\begin{figure}[tbh]
{\includegraphics[width=0.9\columnwidth]{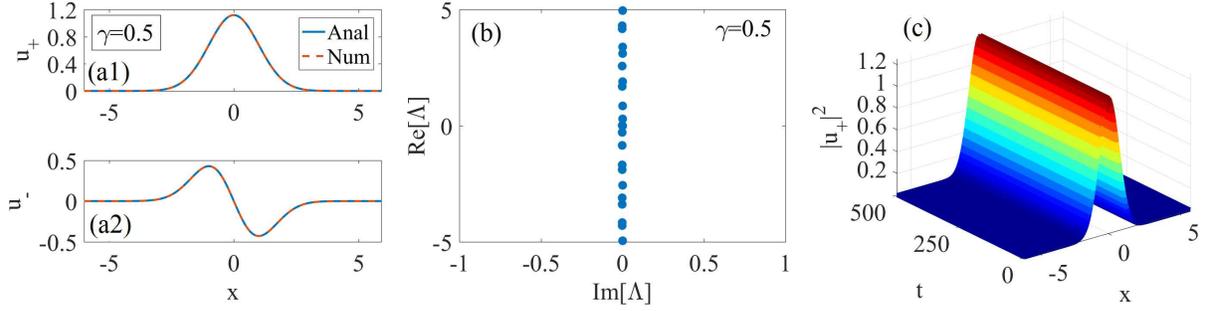}}
\caption{(a1,a2) The comparison between the analytical and numerical results
(the blue solid and red dashed curves) for the 1D wave function with $%
\protect\gamma =0.5$. (b) The spectrum of stability eigenvalues $\Lambda $
for the exact solution from panels (a1,a2). (c) Direct simulations of the
evolution of this soliton. Here the analytical wave functions are given by
Eqs. (\protect\ref{twoansatz}), (\protect\ref{uu}), (\protect\ref{APoneD})
and (\protect\ref{ANoneD}), while $\protect\lambda $ and $N$ are defined by
Eqs. (\protect\ref{lambda1D}) and (\protect\ref{totalnorm1D}), respectively.
}
\label{stable1D}
\end{figure}

Comparison between the analytical solution and numerical results, produced
by the imaginary-time integration of Eq. (\ref{1D}), for $\mu $ and $F_{-}$
in the regions of $0\leq \gamma <1$ and $\gamma >3$ is displayed in Fig. \ref%
{NumAna1D}. Similar to the situation in the 2D system, cf. Fig. \ref{NumAna}%
, in the region of $0\leq \gamma <1$ the analytical solution exactly
produces the ground state, while at $\gamma >3$, the analytical solution
corresponds to some excited state, the ground state being strongly
different. The corresponding comparison of the wave functions, as produced
by the numerical and exact solutions in the regions of $0\leq \gamma <1$ and
$\gamma >3$, are displayed in Figs. \ref{stable1D}(a1,a2) and \ref%
{unstable1D}(a1,a2), respectively.
\begin{figure}[tbh]
{\includegraphics[width=0.6\columnwidth]{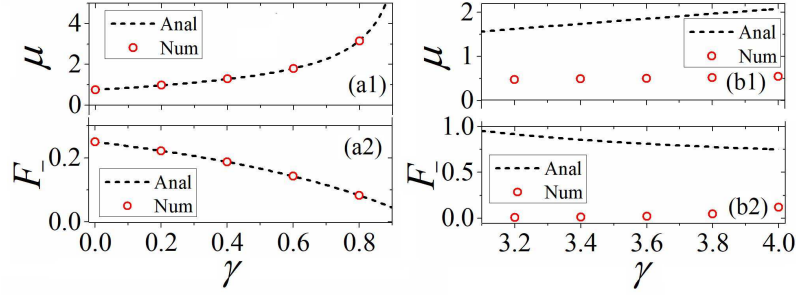}}
\caption{(a1,a2) The comparison between the exact analytical result (the
black dashed curve) and numerical findings, produced by the imaginary-time
simulations of the 1D system (red circles) for $\protect\mu $ and $F_{-}$ in
the region of $0\leq \protect\gamma <1$. (b1,b2) The same at $\protect\gamma %
>3$. Parameters $\protect\lambda $ (the strength of the SO coupling) and $N$
(the total norm) are defined by Eqs. (\protect\ref{lambda1D}) and (\protect
\ref{totalnorm1D}), respectively. The analytical expressions for $\protect%
\mu (\protect\gamma )$ and $F_{-}(\protect\gamma )$ are given by Eqs. (%
\protect\ref{mu1D}) and (\protect\ref{FN1D}), respectively. }
\label{NumAna1D}
\end{figure}

\begin{figure}[tbh]
{\includegraphics[width=0.9\columnwidth]{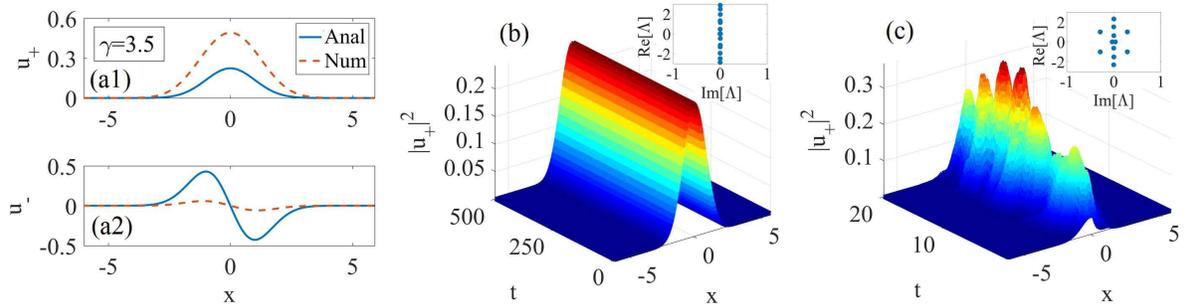}}
\caption{(a1,a2) Comparison between the analytical and numerically found
(the blue solid and red dashed curves, respectively) for the 1D soliton wave
function at $\protect\gamma =3.5$. (b,c) Direct simulations to the evolution
initiated by the numerical and analytical solution, respectively. Insets
arespectra of stability eigenvalues $\Lambda $ for them. Here the analytical
wave functions are defined by Eqs. (\protect\ref{twoansatz}), (\protect\ref%
{uu}), (\protect\ref{APoneD}) and (\protect\ref{ANoneD}), while $\protect%
\lambda $ and $N$ are defined by Eqs. (\protect\ref{lambda1D}) and (\protect
\ref{totalnorm1D}), respectively. }
\label{unstable1D}
\end{figure}

The stability of the analytical and numerical solutions was verified by
numerical computation of the respective eigenvalues and through direct
simulations of Eq. (\ref{1D}). Also similar to the 2D case, the simulations
show that the analytical exact solution, along with its numerical
counterpart, are completely stable in the interval of $0\leq \gamma <1$, see
a typical example for $\gamma =0.5$ in Fig. \ref{stable1D}. On the other
hand, in the region of $\gamma >3$, the numerical solution is stable, while
the exact one is not. Typical examples of the stable and unstable evolution
for $\gamma =3.5$ are displayed in Fig. \ref{unstable1D}.
\begin{figure}[tbh]
{\includegraphics[width=0.6\columnwidth]{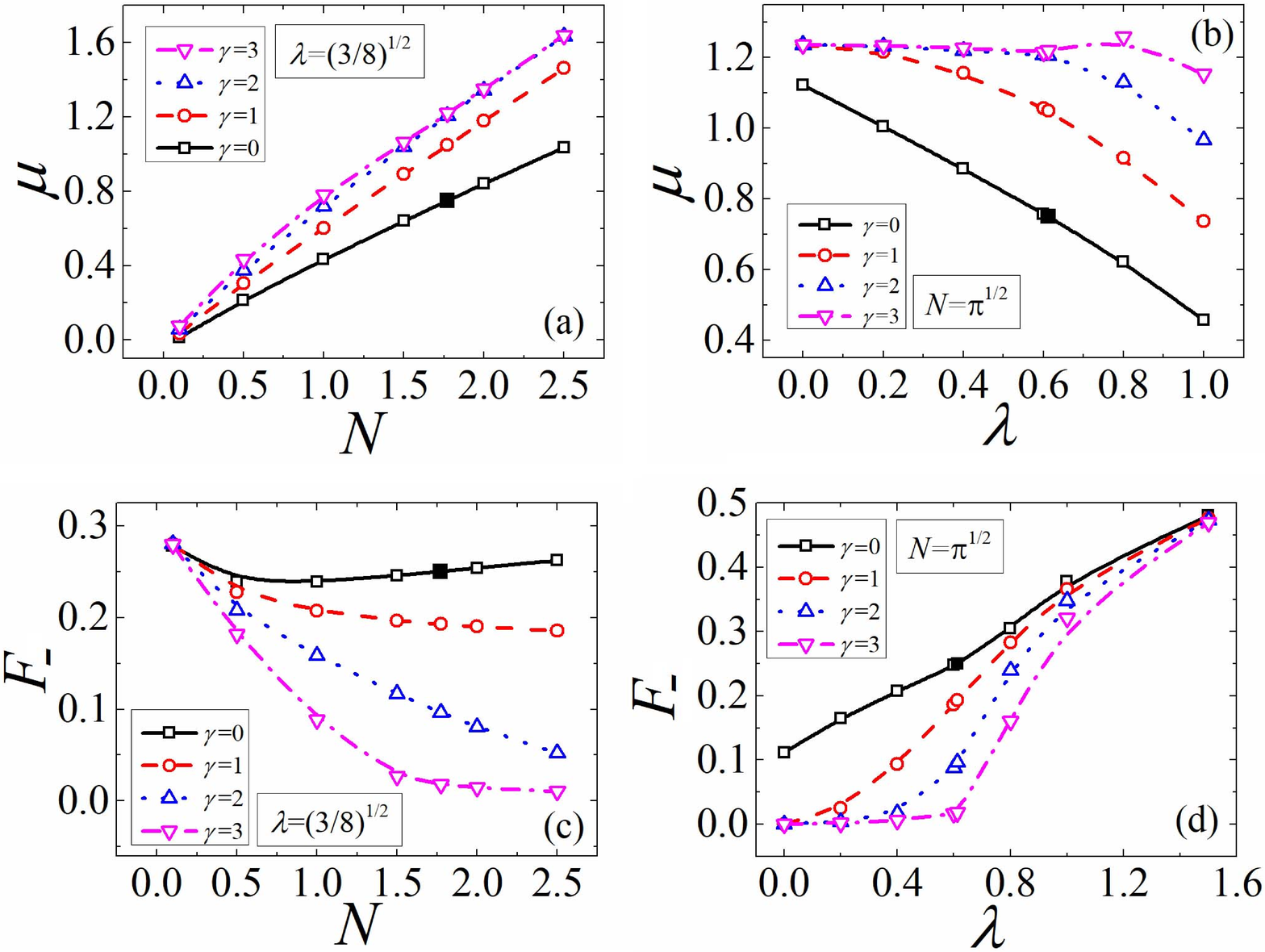}}
\caption{The chemical potential of the numerically found family of 1D
semi-vortices, $\protect\mu $, versus the total\ norm, $N$, and the
SO-coupling strength, $\protect\lambda $, for different values of the
cross/self interaction ratio, $\protect\gamma =0$, $1$, $2$, $3$,
respectively. (c,d) The norm share in the vortical component, $F_{-}$ [see
Eq. (\protect\ref{N_})], versus $N$ and $\protect\lambda $ for $\protect%
\gamma =0$, $1$, $2$, and $3$, respectively. In panels (a,c) and (b,d), $%
\protect\lambda =\protect\sqrt(3/8)$ and $N=\protect\sqrt(pi)$ is fixed,
severally. The black solid squares in the panels represent the exact
solution, given by Eqs. (\protect\ref{uu})-(\protect\ref{FN1D}), with $%
\protect\gamma=0$. }
\label{muFN1D}
\end{figure}

\begin{figure}[tbh]
{\includegraphics[width=0.9\columnwidth]{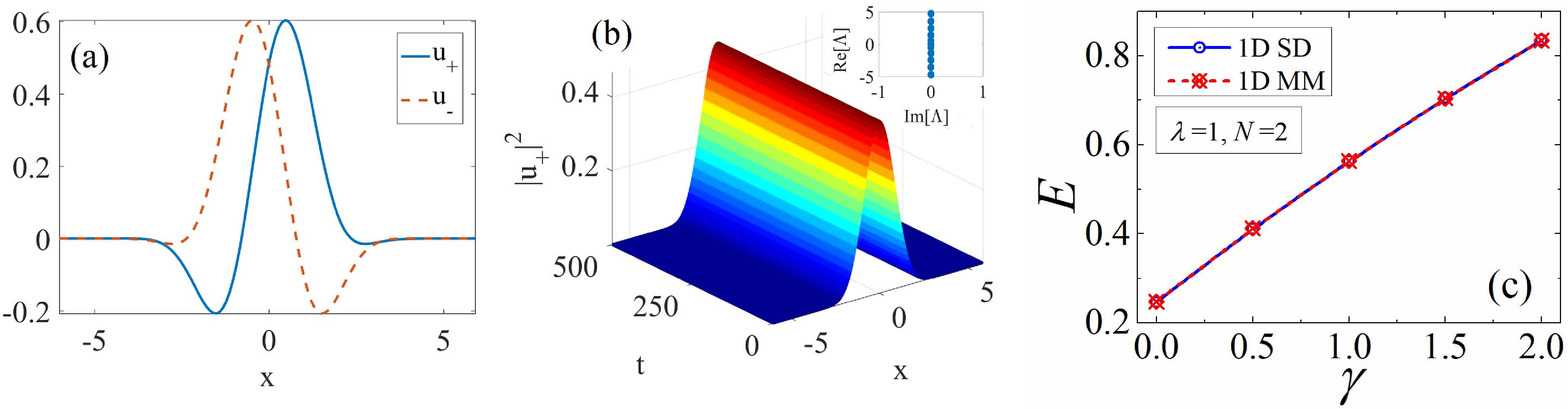}}
\caption{(a,b)A typical example of numerically found stable 1D mixed-mode
solitons with $(N,\protect\gamma,\protect\lambda)=(2,0,1)$. (c) Energies of
the 1D semi-dipole and MM solitons versus $\protect\gamma $ for fixed $N=2$
and $\protect\lambda =1$. }
\label{MM1D}
\end{figure}

Following the patterns of the analysis presented in subsection B of Section
III, families of generic semi-dipole states can be found by means of the
imaginary-time method, applied to Eq. (\ref{1D}) with the input taken as per
Eq. (\ref{uu}). Similar to Fig. \ref{SVmuFN}, we characterize the
semi-dipole families by plotting $\mu$ and $F_{-}$ versus $N$ and $\lambda$
in the Fig. \ref{muFN1D}. The exact analytical solution, given by Eqs. (\ref%
{uu}) and (\ref{lambda1D})-(\ref{FN1D}), is included in the figures.

Further, the 1D version of MMs can also be found in the numerical form,
starting from ansatz
\begin{equation}
u_{\pm }=\left( A_{1}\pm A_{2}x\right) \exp \left( -x^{2}/2\right).
\end{equation}%
A typical example of the stable 1D MM for $\gamma=0$ is displayed in Fig. %
\ref{MM1D}. The $E(\gamma)$ curves for the 1D semi-dipoles and MMs are shown
in the Fig. \ref{MM1D}(c), demonstrate that these two species of 1D solitons
are almost degenerate, in terms of the energy.

\section{Conclusion}

The objective of this work is to construct 2D solitons in the SO
(spin-orbit)-coupled two-component BECs, with the contact repulsive
interactions whose local strength grows from the center to periphery
sufficiently fast. With the anti-Gaussian modulation profile, exact
analytical solutions of the SV\ (semi-vortex) type have been found. These
solutions are chiefly controlled by the relative strength of the cross
repulsion, $\gamma $. The exact solutions exist in the regions of $\gamma <1$
and $\gamma >2$. Numerical results demonstrate that exact solutions are
stable ground states at $\gamma <1$, while at $\gamma >2$ they are unstable
excited modes, the ground state being a different solution, found in the
numerical form. Other types of 2D solitons, \textit{viz}., fundamental MMs
(mixed modes), and excited state of SVs and MMs are found numerically, as
well as the full family of SVs into which the exact analytical solutions are
included as particular ones. All the fundamental solutions of both the SV
and MM types are completely stable. They are identified as the ground state
at $\gamma >1$ and $\gamma <1$, respectively, while their excited states,
produced by adding vorticity $S$ to both components, are partly stable up to
$S=5$ and $2$, for the SVs and MMs, respectively. The 1D reduction of the
system was considered too. In particular, exact states, in the form of
semi-dipoles, which are 1D counterparts of the SVs, were found at $\gamma <1$%
. Families of generic 1D solitons, of the semi-dipole and MM types, have
been found in the numerical form. These two types of 1D solitons are close
to being mutually degenerate, as they have almost equal energies.

The present analysis can be further extended. First, a natural possibility
is to explore the SO-coupled system with a mixture of Rashba and Dresselhaus
coupling terms. Next, one can consider the limit case of strong SO coupling,
which makes it possible to neglect the kinetic-energy terms in Eq. (\ref%
{basicEq}), thus replacing it by a nonlinear Dirac/Weyl model. Finally, a
challenging option is to seek extension of the current setting to the 3D
geometry.

\begin{acknowledgments}
This work was supported, in part, by NNSFC (China) through Grant No.
11575063, 61471123, 61575041, by the joint program in physics between NSF
and Binational (US-Israel) Science Foundation through project No. 2015616,
by the Israel Science Foundation (project No. 1287/17), and by the Natural
Science Foundation of Guangdong Province, through Grant No. 2015A030313639.
B.A.M. appreciates a foreign-expert grant of the Guangdong province (China),
and a Ding-Ying professorship, provided by the South China Agricultural
University (Guangzhou) at its College of Electronic Engineering.
\end{acknowledgments}

\end{document}